\documentclass[%
 reprint,
 amsmath,amssymb,
 aps,
]{revtex4-2}

\usepackage{graphicx}
\usepackage{dcolumn}
\usepackage{bm}

\begin{document}

\preprint{APS/123-QED}

\title{Solution of one-dimensional Kondo lattice model, ground state calculation}

\author{Igor N.Karnaukhov}
\affiliation{G.V. Kurdyumov Institute for Metal Physics, 36 Vernadsky Boulevard, 03142 Kiev, Ukraine\\
Department of Physics, Bar-Ilan University, Ramat-Gan 52900, Israel}

\date{\today}

\begin{abstract}
The ground state of the Kondo chain is calculated taking into account the formation of local singlet states of electrons and moments.
Singlets are entangled local states of electrons and moments arranged chaotically and varying in time. Two-particle scattering matrix of electrons forming singlets is calculated using the Bethe Ansatz. It is shown that electrons do not hybridize with local moments, and a lattice with a double cell is not formed. In the Kondo insulator a charge gap is calculated for an arbitrary value of the exchange integral. In the case of strong interaction the gap is determined by the single-particle energy of the singlet, for weak interaction - by correlations (the gap is proportional to the square of the exchange integral).
\end{abstract}

\maketitle


\section{\label{sec:level1}Introduction}

The Kondo and Anderson lattice models describe coexistence of interacting delocalized and localized states, and  formation of the Kondo insulator at half-filling.   In contrast to the Anderson chain, where $d$-electrons form flat bands, in the Kondo chain localized electrons are represented by magnetic moments regularly located on lattice sites.
Exact solutions of the Kondo problem and  of the Anderson model allowed  to study the behavior of the electron liquid in the case of weak interaction in the continuum approach  \cite{1a,2a}. In the weak interaction limit, the Kondo temperature (corresponding energy scale) is exponentially small in the antiferromagnetic exchange interaction constant.

The Kondo chain is intensively studied as it concerns the Kondo effect, heavy fermions,  various magnetic phase states. The one-dimensional Kondo lattice also allows a better understanding of the Kondo insulator  (see reviews \cite{b1,b2,b3}).
It is well known that the scattering of electrons on a local moment with spin flip dominates and leads to formation of a singlet state of electrons and local moment in the Kondo problem. Most likely, it is this scattering that also forms the Kondo insulator, so it must be explicitly taken into account in  calculations.
Despite the rather large scientific interest in  the Kondo insulator problem \cite{K1,K2,K4,f,f1}, we do not even know which ground state corresponds to this phase.  
There are still open questions related to the nature of the Kondo insulator: whether there is a large Fermi surface, what are the values of the spin and charge gaps, and most importantly, what is the ground state of the electron liquid \cite{2,3,4}.  The concept of the Luttinger liquid which is realized in one-dimensional interacting electron models is most likely the most productive for explaining the phase diagram of the Kondo chain  \cite{5}.
One cannot speak unambiguously about the behavior of the electron liquid in the Kondo insulator even in the case of weak coupling, since the singlet formation process is nonperturbative. A rather large gap in the spectrum of the  Kondo insulator 700 K in $FeSi$ \cite {f1} cannot be explained in a weak coupling limit.  At that time, the Kondo insulators containing rare-earth or actinide elements $YB_{12}$, $CeNiSn$, $Ce_3Bi_4Pt_3$ have a gap in the spectrum of quasiparticles excitations of the order of a few $meV$. The one-dimensional model does not describe real compounds, nevertheless it gives an opportunity to analyze the physical nature of the phenomenon. In this paper, the emphasis is not on explaining the experimental data, but on understanding the realization of the ground state of an electron liquid in the Kondo insulator.

The paper is devoted to the study of the most interesting state, which is realized in the Kondo lattice when it is half-filled, we will talk about the Kondo insulator. A rather stable representation of the ground state of the Kondo insulator is related to the doubling of the cell \cite{b1,f}. It is clear that in this case as a rule there is a gap in the spectrum of charge excitations even in the weak coupling limit.   
In this paper we propose a new approach to the study of the Kondo insulator. Singlet states of electrons and local moments (with minimum energy \cite{b1}) are formed chaotically when electrons move along the chain.  Taking into account the lowest-energy bound states of electrons and moments (local singlet states), the scattering of electrons is calculated using the Ansatz Bethe. 
Electrons and moments form singlet states on the same lattice sites, and at half filling 
the ground state of the Kondo insulator is formed by these local singlet states. Sufficiently productive is the notion of $Z_2$-field, which allows us to to take into account the behavior of the band states in random static field formed by the states localized in the lattice sites \cite{AK}.

\section{The Kondo chain}
\label{sec:1}
We will start with the  Hamiltonian of the spin-$\frac{1}{2}$ Kondo chain
\begin{eqnarray}
 &&{\cal H}= - \sum_{<i,j>}\sum_{\sigma=\uparrow,\downarrow}
c^\dagger_{i \sigma} c_{j \sigma}+ \sum_{j=1}^N [ J_\Vert s^z_j S^z_j+\nonumber\\&&\frac{1}{2}J_\bot( s^+_{j}S^-_j+s^-_{j}S^+_j)],
\end{eqnarray}
where
$c^\dagger_{j \sigma}$ and $c_{j \sigma}$ are the fermion operators  for a  electron determined on a lattice site $j$  with spin $\sigma =\uparrow,\downarrow$ (or $\pm $), the hopping integral between the nearest-neighbor lattice sites is equal to one, the spin operators for the electrons are defined  as  $s^z_j=\frac{1}{2}(n_{j \uparrow}-n_{j \downarrow})$,  $s^+_{j}=c^\dagger_{j \uparrow} c_{j \downarrow}$, $s^-_{j}=c^\dagger_{j \downarrow} c_{j \uparrow}$,  $n_{j\sigma}=c^\dagger_{j \sigma} c_{j \sigma}$,
are the density operators, $\textbf{S}_j$ are the Pauli matrices, which define the spin-$\frac{1}{2}$  operator on a lattice site $j$, $J_\Vert > 0$ and $J_\bot > 0$ are the magnitudes of an anisotropic antiferromagnetic  exchange  interaction,  N is the total number of lattice sites.
 
Using the definition of the $S$-spin operators via the one electron $a$-Fermi operators let us redetermine the interaction term in (1) as follows 
\begin{eqnarray}
 &&
\sum_{j=1}^N [ \frac{1}{2}J_\Vert (n_{j \uparrow}-n_{j \downarrow})(m_{j \uparrow}-m_{j \downarrow})+
\nonumber\\
&&
J_\bot ( c^\dagger_{j \uparrow} c_{j \downarrow} a^\dagger_{j \downarrow} a_{j \uparrow}+ c^\dagger_{j \downarrow} c_{j \uparrow} a^\dagger_{j \uparrow} a_{j \downarrow})],
\end{eqnarray}
{where $m_{j\sigma}=a^\dagger_{j \sigma} a_{j \sigma}$.
The energy of the electron at the site will be counted from the energy $\varepsilon_g<0$ with the introduction of an additional term $\varepsilon_g\sum_{j}\sum_\sigma m_{j\sigma}$ in (2). The state of two particles $m_{j\uparrow}m_{j\downarrow}$ is forbidden and is not considered.}
Scattering with spin flipping results in the formation of the ground state of the electron liquid in the Kondo problem and the Kondo lattice; in (1),(2) this process is accounted for by the exchange interaction \cite{K1,K2,K4}.

\section{\label{sec:level1}Solution of the problem}

Spin-$\frac{1}{2}$ moments located at the lattice sites form a static $Z_2$ field in which electrons move.
The configuration of this field corresponding to the minimum energy determines the ground state of the electron liquid. The configuration corresponding to the ground state does not break the translational symmetry of the lattice, a free configuration of $Z_2$-field \cite{Lieb} is realized  in the gapless Majorana spin liquid in the Kitaev model \cite{AK}.
 In contrast to the previous works \cite{K1,K2,K4}, where the Hubbard-Stratonovich transformation was used to define the   $Z_2$-field, we will proceed from the natural assumption that the free configuration of the field corresponds to the ground state of the electron liquid. In the mean-field approximation,  $Z_2$-field, which is determined by the interaction between electrons and local moments, leads to the Kondo insulator with the double-cell lattice   \cite{K1,K2,K4,f}. It should be noted that in the Hubbard model we have the same behavior, namely, in the mean-field approach the Mott insulator corresponds to a lattice with a doubled cell, while at the same time the exact solution of the Hubbard chain does not correspond to cell doubling \cite{LW}.
Let us define this field configuration and at the same time consider the motion and scattering of electrons

The moments located at each lattice site form an effective field with which  electrons interact due to the exchange interaction (1).
First of all, let us calculate the energies of the local states of an electron taking into account only the exchange interaction in (1).
Solutions for one-electron wave functions   $\psi({j},\sigma,\sigma') c^\dagger_{{j} \sigma}a^\dagger_{j\sigma'}$ determine the energies of the local states  $\epsilon$. $\psi$-amplitudes satisfy the following equations:
\begin{eqnarray}
&& (\epsilon-\frac{1}{2}J_\Vert)\psi({j}, \sigma,\sigma) =0, \nonumber\\
&&(\epsilon+\frac{1}{2}J_\Vert)\psi({j},\sigma,-\sigma) -J_\bot \psi({j},-\sigma, \sigma)=0, \nonumber\\
&&(\epsilon+\frac{1}{2}J_\Vert)\psi({j},-\sigma,\sigma) -J_\bot \psi({j},\sigma,-\sigma)=0,
\end{eqnarray}

The wave functions with $\psi({j},\sigma,-\sigma)\neq 0$ determine the local states  of  electron and moment. A minimal energy, equal to $\epsilon =-c$ ( $c=J_\bot+\frac{1}{2}J_\Vert$), corresponds to singlet state with $\psi({j},\sigma,-\sigma)=-\psi({j},-\sigma,\sigma)$.
Energies  $\epsilon =-\frac{1}{2}J_\Vert +J_\bot$ and $\epsilon =\frac{1}{2}J_\Vert $  correspond to other solutions $\psi({j}, \sigma,-\sigma) =\psi({j}, -\sigma,\sigma) $ and $\psi (j,\sigma,\sigma)$.  Singlets defined by the operators $X_j=c^\dagger_{j \uparrow} a^\dagger_{j\downarrow}-c^\dagger_{j \downarrow}a^\dagger_{j \uparrow}$, whose energies are $-c$, form the ground state.

Thus two electrons with different spins and a local moment form local singlets on  lattice sites, when electrons moves along the chain, in our case we are talking about the vacuum configuration, which corresponds to the ground state of an electron liquid. Note that we consider only the case $J_\bot >0$, $J_\Vert >0$ since in this case the energy of $X$-singlet is minimal. 

We introduce  the operators $q_j=a^\dagger_{j\uparrow}+a^\dagger_{j\downarrow}$, these operators determine vacuum configuration or a field in which the electrons move. On the one hand, the exchange interaction in the Hamiltonian (2) leads to the formation of local $X$-singlets, on the other hand, the $q$-vacuum also responds to the formation of the singlets.
The spin subsystem adjusts to the states of the electrons at each lattice site, forming singlets. Singlets mean entangled local states of electrons and moments arranged chaotically. This process is dynamic, its dynamics is determined by the motion of electrons, let us consider it in more detail. We consider the equation for the wave function $\psi(j)(c^\dagger_{j,\uparrow}+c^\dagger_{j,\downarrow}) q_j$ in the chain with $q_j$ vacuum configuration
(two electrons with different spins form a singlet).
Note that at motion of the $(c^\dagger_{j,\uparrow}+c^\dagger_{j,\downarrow})$-operator the vacuum state at each lattice site does not change. We can consider solutions for the wave function with corresponds to following equation
\begin{eqnarray}
&&[\epsilon (k)+c]\psi({j})+ \sum_{\textbf{1}}\psi(j+\textbf{1})=0,
\end{eqnarray}
where it is summed over the nearest lattice sites, $ \epsilon (k)=-2\cos k -c$, $k$  is  the  wave vector of an electron.
 
Eq (4) defines the 'effective motion' of $X-$singlet, when electron with spin-$\uparrow$ (or $\downarrow$)
is moved along the chain, here is a direct correspondence between moving electrons and singlets.
The solution $\psi({j},\sigma,-\sigma)=-\psi({j},-\sigma,\sigma)$  (see Eq (3)) is valid at each lattice site, and the high-energy states of electrons are not taken into account.
The calculations also take into account that an electron with momentum $k$ forms a singlet only on one lattice site, as a result of which the vacuum configuration does not change on lattice site $j$ when the electrons move to site $j+\textbf{1}$. 

The two-particle wave function of the electrons describes the scattering of electrons defined on $q$-vacuum configuration.
Due to Eq (4) the equation for the wave function $\psi(j_1,j_2)$, defined on $q$-vacuum, satisfies the following equation at $j_1\neq j_2$
\begin{eqnarray}
&&[\epsilon (k_1)+ \epsilon (k_2)+2c]\psi(j_1,j_2)+\nonumber \\&&
 \sum_{\textbf{1}}[\psi(j_1+\textbf{1},j_2) + 
 \psi(j_1,j_2+\textbf{1})]=0.
 \end{eqnarray}
The equation for the wave function $\psi(j,j)$ at $j_1= j_2=j$ has the following form
\begin{eqnarray}
&&[\epsilon (k_1)+ \epsilon (k_2)]\psi(j,j) +\sum_{\textbf{1}}[\psi(j+\textbf{1},j) +
  \psi(j,j+\textbf{1}) ] =0.\nonumber \\
\end{eqnarray}
 Two electrons located on the same lattice site do not interact with the local moment through exchange interaction in (1),(2), which is taken into account in Eq (6).
Due to the fact that the two-electron $X_j$-operator is a boson, the function $\psi(j_1,j_2)$ is coordinate-symmetric, it determines the scattering processes of electrons that move in the $q_j$-vacuum and scatter on the on-site potential.

Let's examine the equation for the wave function at $j_1=j_2=j$ in more detail.  
In Eq (6), a local operator 
$c^\dagger_{j,\downarrow}c^\dagger_{j,\uparrow}(a^\dagger_{j,\downarrow}+a^\dagger_{j,\uparrow})$ or 
$c^\dagger_{j,\downarrow}c^\dagger_{j,\uparrow}q_j$
corresponds to amplitude $\psi(j,j)$.
At transition of one electron operators to neighboring sites, we obtain the following operators
$c^\dagger_{j\pm 1,\sigma}q_{j\pm 1}c^\dagger_{j,-\sigma}q_j$. 
The wave functions, defined on the operators are $\psi(j\pm 1)\psi(j,)$ (see Eqs (3),(6)).
The wave function describes the scattering of electrons on on-site  effective potential equal to 2c. 
The solution for the wave function is defined by the Ansatz Bethe, the scattering matrix is defined as
\begin{eqnarray}
{\cal S}_{1,2}=\frac{\sin k_1-\sin k_2- i c }{\sin k_1-\sin k_2 +i c }.
\end{eqnarray}
The two-particle scattering matrix (7) describes the scattering of two electrons forming singlets.

The solution of the problem reduces to solving a set of ${N_e}$ coupled algebraic equations for the $N_e$
quasi-momenta $k_j$,  the Bethe equations are determined as
\begin{eqnarray}
&&\exp( ik_j N)=\prod_{i\neq j}^{N_e}\frac{\sin k_j -\sin k_i +i c}{\sin k_j -\sin k_i -i c},
\end{eqnarray}
where $N_e$ is the total number of electrons.

In the thermodynamic limit ($N\to \infty$, with $\rho =N_e /N$ is fixed) the structure of solutions of
the Bethe  equations (8) for the ground-state system includes real quasi-momenta.  The ground state of the model is the
Fermi sea, characterized by  the distribution function of quasi-momenta  $\rho (k)$. The  'Fermi level' is determined by  $\lambda$ value,
the ground-state configuration corresponds to filling of all states with $-\lambda< k <\lambda$.  According to (8)
the function $\rho (k)$ satisfies the following integral equation
\begin{eqnarray}
&&
\rho (k)=\frac{1}{2 \pi}+ \frac{1}{2\pi}\cos k\int_{-\lambda}^{\lambda}dq \frac{c}{(\sin k -\sin q)^2+c^2}\rho (q).\nonumber\\
\end{eqnarray}
The density of the ground state energy for $\rho\leq 1$  is equal  to
\begin{eqnarray}
&&
E(\lambda)=-2\int_{-\lambda}^{\lambda} dk \cos k \rho(k) -c\rho,
\end{eqnarray}
where $\rho=\int_{-\lambda}^{\lambda} dk \rho(k) $.

For the dressed energy function  $\varepsilon (k)$  we have the following equation
\begin{eqnarray}
&&
\varepsilon (k)=-2\cos k+ \frac{1}{2\pi}\int_{-\lambda}^{\lambda}dq \cos q \frac{c}{(\sin k -\sin q)^2+c^2}\varepsilon (q).\nonumber\\
\end{eqnarray}
Solution $\lambda=\lambda_0=\pi$ corresponds to half-filling $\rho=1$ and  solution for $\rho (k)$ has the following form
\begin{eqnarray}
&&
\rho (k)=\frac{1}{2 \pi}+ \frac{1}{4\pi^2}\cos k\int_{\lambda_0}dq \frac{c}{(\sin k -\sin q)^2+c^2}.\nonumber\\
\end{eqnarray}
The substitution of (12) into (10) yields the ground state energy at half-filling
\begin{eqnarray}
&&
E_0=- \frac{1}{2\pi^2}\int_{\lambda_0} dk \cos^2 k \int_{\lambda_0}dq \frac{c}{(\sin k -\sin q)^2+c^2}-c\rho.
\nonumber \\
\end{eqnarray}

\subsection{Charge gap calculation}
Let us consider two states of the system with differen filling: the first at half filling with $N_e=N$ singlets, the second with $N_e=N+1$, $N$  singlets and an electron, by an additional electron. All processes with virtual hoppings of singlets are reduced to the permutations of singlets, in which case the energies of the initial and final states are identical.
We will take into account the excitations of the ground state due to the simplest virtual hoppings of singlets described by the scattering matrix (7). The quasimomentum $k_i$ corresponds to two electrons with different spins that form a singlet, so the function $\psi(j)$ (4) is spin-independent. In this case when one electron is added we have a 'half-electron' (with density operator  $n_{j\sigma}=\frac{1}{2}$) which corresponds Bete equations (8). The extra electron destroys one singlet, since  are two electrons  located at lattice site destroy a singlet at the same site.  The new state contains $N-1$ singlets, one electron into a full filled subband (with the Fermi energy) and additional a 'half-electron' with moments $k_0$ energy $-\cos k_0$, which moves along the chain. The motion of this electron along the chain is determined by Eqs (3), the energy of such state is larger by the value of c, since the two-electron state at a lattice site does not interact with the local moment. The equation for the distribution function $\rho (k)$ for $N-1$  singlets is transformed to
\begin{eqnarray}
&&
\rho (k)=\frac{1}{2 \pi}+ \frac{1}{2\pi}\cos k\int_{\lambda_0}dq \frac{c}{(\sin k -\sin q)^2+c^2}\rho (q)-\nonumber\\&&
\frac{1}{N}\delta (k-k_0).
\end{eqnarray}
Introducing the definition $\rho (k)=\rho_0 (k)+\frac{1}{N}\delta \rho (k)$, we redefine equation (14) in terms of $\delta \rho (k)$, it has the following form
\begin{eqnarray}
&&
\delta \rho (k)=-\frac{\cos k}{2\pi} \frac{c}{(\sin k -\sin k_0)^2+c^2}.
\end{eqnarray}
Formulas  (10), (15) make it possible to calculate the energy difference corresponding to these states $\frac{1}{N}\delta E=E(\lambda)-E_0$
\begin{eqnarray}
&&
 \delta E=c - \cos k_0 -\int_{\lambda_0} \cos k \delta \rho (k) dk.
\end{eqnarray}
A minimal value of $\delta E$ as function of momenta  $k_0$  determines the gap  $\Delta$ in the Kondo insulator. The solution $k_0=0$ corresponds to the value of $\Delta$
\begin{eqnarray}
&&\Delta = c- 1+\frac{1}{2\pi} \int_{\lambda_0}dk  \cos^2 k  \frac{c}{\sin^2 k +c^2}.
\end{eqnarray}
The equation for $\Delta$ is similar to that in the Hubbard chain \cite{LW}, the kernel of the integral equations for the momentum distribution function is different.  In the strong interaction limit $c\to \infty$ $\Delta =c-1$ the summand associated with correlation effects is annihilated, $\delta \rho (k)  \sim \frac {1}{c}$. In the $c\to 0$ limit $\Delta \approx\frac{1}{2}c^2$, this summand dominates in the case of  $c<<1$. Numerical calculations the value of the gap is shown in Fig.1.

The value of the gap differs from such in the one-dimensional Hubbard model \cite{LW}. The electrons at each lattice site form singlets, and we are talking about bosons rather than fermions.  The scattering matrix in this case has the traditional form for two-particle scattering of bosons. The gap in the quasiparticle spectrum is determined by two-particle  scattering of singlets; in the case of weak interaction it is proportional to the square of the exchange integral (or the constant of the Ruderman–Kittel–Kasuya–Yosida (RKKY) interaction), while the antiferromagnetic state is not realized. 

The charge and spin gaps have different energy scales \cite{b1}. In the case of a weak exchange interaction the spin gap is exponentially small \cite{b1,f,b4}, the charge gap, calculated by the density matrix method
\cite{b4}, is proportional to $\frac{1}{2}J$ ($J$ is the exchange integral). Unfortunately, numerical calculations of the gaps were obtained for chains of small length with $N\leq 8$ \cite{b1} and we cannot say about the exact behavior of the charge gap in the case of weak interaction. It should be noted that in the Kondo problem there are two energy scales: the Kondo temperature and the energy of the RKKY-interaction. It is not excluded that the first scale determines the magnitude of the spin gap, the second one the charge gap.

In the Kondo insulator the behavior of the Kondo chain is similar to the Kondo problem at small screening  radius of local moment (in the Kondo lattice there is local screening). The nature of screening by conduction electrons of a local moment in the Kondo problem and the Kondo lattice are quite different. In the Kondo problem the electron cloud polarizes the local moment of the impurity, in the Kondo lattice local singlets (conduction electrons and moments) are formed, the scattering of which determines the behavior of the Kondo insulator.

\begin{figure}
\centering{\leavevmode}
\begin{minipage}[h]{1\linewidth}
\resizebox{0.85\hsize}{!}{\includegraphics*{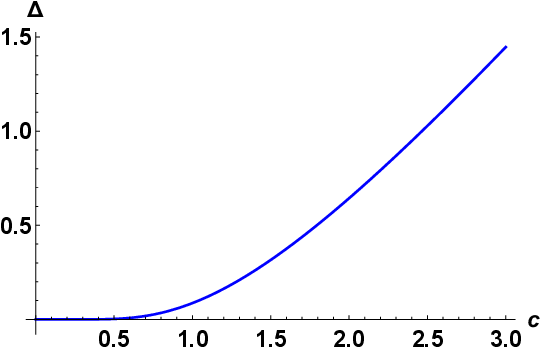}} 
\end{minipage}
\caption{(Color online) The charge gap $\Delta $ in the Kondo insulator as function of the exchange integral $ c=\frac{J_\Vert}{2}+J_\bot$.
  }
\label{fig:1}
\end{figure}

\section{Conclusion}
Within the framework of the spin-$\frac{1}{2}$ model of the Kondo chain, the ground state of the electron liquid is studied.  In the Kondo chain, electrons and local moments form local singlets, which are arranged chaotically \cite{K4,f}. Electrons and local moments do not hybridize, no large Fermi surface is formed (which is consistent with \cite{K4,f}), cell doubling is not realized (not consistent with \cite{K4,f}). The charge gap in the Kondo insulator is determined by correlations in the weak interaction and the one-particle interaction for the strong interaction. In the case of a weak interaction limit the charge gap is proportional to the square of the exchange integral. In the strong interaction limit the gap calculations agree with those obtained in the mean-field approximation \cite{K4}.
In this paper, a new mechanism for the formation of the ground state of the Kondo insulator is proposed.  The calculations explicitly take into account the formation of singlets at chaotic motion of electrons and the scattering of electrons during their motion.

\begin{acknowledgments}
The author is grateful to Bar-Ilan University for support under the Bar-Ilan  University International School program, and thanks R. Berkovits and D. Golosov for  hospitality and  discussions.

\end{acknowledgments}

\end{document}